\documentclass[12pt,tightenlines,eqsecnum,floats,showpacs,nofootinbib,amsmath,amssymb,aps,prd,superscriptaddress]{revtex4-1}

\usepackage{graphicx,color}
\usepackage{amsmath,amssymb}
\usepackage{hyperref}

\begin{document}
	
\title{Modified FRW cosmologies arising from states of the hybrid quantum Gowdy model}

\author{Beatriz Elizaga Navascu\'es}
\email{beatriz.elizaga@iem.cfmac.csic.es}
\affiliation{Instituto de Estructura de la Materia, IEM-CSIC, Serrano 121, 28006 Madrid, Spain}
\author{Mercedes Mart\'in-Benito}
\email{mmartin@hef.ru.nl}
\affiliation{Radboud University Nijmegen, Institute for Mathematics, Astrophysics and Particle Physics, Heyendaalseweg 135, NL-6525 AJ Nijmegen, The Netherlands}
\author{Guillermo A. Mena Marug\'an} \email{mena@iem.cfmac.csic.es}
\affiliation{Instituto de Estructura de la Materia, IEM-CSIC, Serrano 121, 28006 Madrid, Spain}
	
\begin{abstract}

We construct approximate solutions of the hybrid quantum Gowdy cosmology with three-torus topology, linear polarization, and local rotational symmetry, in the presence of a massless scalar field. More specifically, we determine some families of states for which the complicated inhomogeneous and anisotropic Hamiltonian constraint operator of the Gowdy model is approximated by a much simpler one. Our quantum states follow the dynamics governed by this simpler constraint, while being at the same time also approximate solutions of the full Gowdy model. This is so thanks to the quantum correlations that the considered states present between the isotropic and anisotropic sectors of the model. 
Remarkably, this simpler constraint can be regarded as that of a flat Friedmann-Robertson-Walker universe filled with different kinds of perfect fluids and geometrically corrected by homogeneous and isotropic curvature-like terms. Therefore, our quantum states, which are intrinsically inhomogeneous, admit approximate homogeneous and isotropic effective descriptions similar to those considered in modified theories of gravity.

\end{abstract}
	
	
\maketitle

\section{Introduction}
\label{sec:Intro}

Nowadays, we are witnessing an outstanding progress in observational cosmology, especially owing to the precise measurements of the cosmic microwave background (CMB), which is one of the best known observational windows to the early epochs of the universe \cite{cmb}. Such a breakthrough provides the opportunity for theoretical physicists to test the predictions of their theories about those early stages. In particular, the possibility that we may falsify our hypotheses about the origin of the universe  in this way, makes clear the necessity of providing a quantum theory for the gravitational interaction, as General Relativity suffers a predictability breakdown in this regime: The big-bang singularity.

One of the most promising approaches to accomplish such a quantization of gravity is based on the formalism of loop quantum gravity (LQG) \cite{lqg}. The application of its techniques to describe the quantization of cosmological models leads to the branch of research known as loop quantum cosmology (LQC) \cite{lqc}. LQC has proven to produce some remarkable results in the analysis of homogeneous models. In particular, the initial big-bang singularity is avoided and replaced with a quantum mechanism, called the big-bounce \cite{lqc,bounce}. However, testing the robustness of LQC calls for obtaining physical results from more realistic scenarios, such as inhomogeneous cosmologies. In this context, a hybrid approach has been proposed for the quantization of models of this kind, based on the assumption that the most relevant quantum geometry effects would mainly affect the homogeneous degrees of freedom. Such assumption involves a splitting of the phase space of the system into a homogeneous sector and an inhomogeneous one. Then, a loop quantization is adopted for the homogeneous degrees of freedom (which, in this way, fully retain the genuine quantum features of the space-time), while the inhomogeneous degrees of freedom are treated by means of a more conventional Fock quantization. This quantization strategy was applied for the first time to the case of the Gowdy cosmologies with linear polarization of the gravitational waves and  with the spatial topology of a three-torus, $T^{3}$, achieving a complete quantization of the model \cite{hybrid,hybrid3}. The study of these tractable cosmologies provides the opportunity to develop approximate methods and techniques to solve the complicated dynamics of inhomogeneous systems. These methods could become particularly interesting when it comes to analyzing the dynamical behavior of more realistic models such as Friedmann-Robertson-Walker\footnote{These cosmologies are also called Friedmann-Lema\^itre-Robertson-Walker cosmologies by many authors.} (FRW) geometries with cosmological perturbations, in the context of an inflationary universe. Recently, these systems have drawn a substantial attention within the framework of LQC. They are being analyzed by means of this hybrid approach \cite{inf-hybrid}, as well as employing other related, but different strategies (see \cite{inf-ash} for a quantum treatment of inhomogeneities over a ``dressed metric'' that accounts for a quantum background without any back-reaction, or \cite{effe} for an effective point of view arising from the requirement that the algebra of constraints closes).

In order to pursue further the analysis of inhomogeneous cosmologies in LQC, we keep investigating the dynamics of the hybrid quantum Gowdy model with linear polarization and $T^{3}$-topology. Specifically we will consider the system with local rotational symmetry (LRS), consistent in this model because it possesses two axial Killing vectors which are indistinguishable in principle, and only one direction is anisotropic (namely, the direction of the inhomogeneities, in which the gravitational waves vary). Besides, to include matter in the system, we will consider a minimally coupled massless scalar field with the same symmetries as the metric \cite{hybrid-matter}. The classical phase space of this Gowdy model can then be seen as that of a flat anisotropic LRS Bianchi I model with inhomogeneities (gravitational waves and matter field) propagating in the anisotropic direction. After a partial gauge fixing, the reduced system is still subject to two global constraints: The zero-mode of the Hamiltonian constraint and a momentum constraint for the inhomogeneous fields. The Hamiltonian constraint operator resulting from the hybrid quantization of this model has a rather complicated action on the states of the kinematical Hilbert space. Thus, approximate methods were developed in \cite{hybrid-approx} in order to find families of states that were approximate solutions to this Hamiltonian constraint. These solutions were actually shown to satisfy as well, within certain approximations, a Hamiltonian constraint that could be seen as corresponding to an FRW universe coupled to a homogeneous massless scalar field and with a perfect fluid. In this paper we will further generalize this family of states so that, while still being approximate solutions to the full Gowdy model, in addition they approximately obey the dynamics corresponding to an FRW cosmology with the kind of geometrical corrections that one would expect to find in modified theories of gravity \cite{modified}. 

The Hamiltonian constraint operator of our hybrid Gowdy cosmology involves two troublesome contributions obstructing its resolution. The first of them is an anisotropy term that contains what can be seen as the momentum of the Bianchi I anisotropy variable. It acts as a difference operator, coupling the isotropic part of the homogeneous sector with the anisotropies. The second contribution is a term that couples the interaction between the inhomogeneous modes (which come from both the gravitational waves and the massless scalar field) with the homogeneous sector. The discussion in \cite{hybrid-approx} and the generalization presented here provide quantum states on which the action of these two complicated terms can be approximated and disregarded when compared to the other terms in the Hamiltonian constraint. Specifically, in the homogeneous sector (which is characterized by the volume of the Bianchi I background and the variable accounting for the anisotropy), the dependence of these states on the anisotropy is given by some particular Gaussian-like profiles. Essentially, these profiles are sharply peaked at a large value of the anisotropy variable, while being reasonably centered at a vanishing value of its momentum. The analyses in \cite{hybrid-approx} dealt with the cases where this peak is either a constant or a volume-dependent function, showing that these states provide approximate solutions to the Gowdy model that dynamically behave as those of an FRW cosmology coupled to a homogeneous massless scalar field (when the peak is a constant) and possibly with a perfect fluid (if the peak depends appropriately on the volume), including the case corresponding to a cosmological constant. The extension of this family of states that we address in this paper considers the more general case in which the peak may depend on a rather generic operator of the homogeneous and isotropic geometry, not given necessarily by a function of the homogeneous volume. It will be argued that, by imposing certain restrictions on this operator, these states provide again approximate solutions to the full Gowdy model and, in turn, to the Hamiltonian constraint of an FRW universe coupled to a homogeneous massless scalar field, filled with a perfect fluid (as in \cite{hybrid-approx}), and now also geometrically corrected by homogeneous and isotropic curvature-like terms. As anticipated above, such new types of corrections can be seen as those expected to arise from certain modified theories of gravity, such as $f(R)$ theories \cite{modified}.

It is worth noting that the considered family of quantum states is not peaked at all at homogeneous and isotropic trajectories, but rather the opposite. Nevertheless, in the dynamics they behave as if they were homogeneous and isotropic, as much as the Hamiltonian constraint is concerned, a fact that can be understood as a quantum collective behavior of their anisotropies and inhomogeneities. Let us also comment that this family is not as limited as one might have thought in principle: Just on the contrary, there exists a considerable variety in the geometry operators that may be used in the construction of the mentioned Gaussian-like profiles.

The organization of the paper is as follows. In Sec. \ref{sec:Gowdy} we will revisit the results of the hybrid quantization of the Gowdy model with $T^{3}$-topology, linear polarization, LRS, and minimally coupled to a massless scalar field \cite{hybrid-matter}. In Sec. \ref{sec:app} we will discuss how, by considering the mentioned Gaussian-like states (already introduced in \cite{hybrid-approx}, and generalized here), the complicated Hamiltonian constraint operator can be approximated by a much simpler one when acting on them. In Sec. \ref{sec:sol} we will justify under which conditions these states can be seen as approximate solutions of the Gowdy model that behave dynamically as those corresponding to a modified FRW cosmology, along the lines commented above. In Sec. \ref{conclu} we will conclude by summarizing our results and discussing their possible implications. Finally, we have included two appendices with extra details of the calculations.

\section{Hybrid quantization of the Gowdy model}
\label{sec:Gowdy}

Let us consider the Gowdy $T^{3}$ model with linear polarization, LRS, and minimally coupled to a massless scalar field $\Phi$ with the same symmetries as the metric.
We will denote the three orthogonal spatial coordinates as $\theta$, $\sigma$, and $\delta$, each of them defined on the circle. This choice of coordinates is adapted to the symmetries of the system, so that the two inhomogeneous fields (matter and gravitational waves) have spatial dependence only in e.g. $\theta$. We can then expand these fields in Fourier modes in this coordinate. The reduced phase space resulting from a partial gauge fixing \cite{hybrid} can be split in two sectors: A homogeneous one formed by the zero-modes of the Fourier expansion, which can be identified with the phase space of a LRS Bianchi I system minimally coupled to a homogeneous massless scalar field $\phi$ (the zero-mode of $\Phi$); and an inhomogeneous sector containing the non-zero Fourier modes of both the linearly polarized gravitational waves and the matter field, as well as their canonically conjugate momenta. The whole of this reduced system is subject to two global constraints: A momentum constraint, $C_\theta$, that generates rigid rotations in $\theta$, and the zero-mode of the Hamiltonian constraint, $C_\text{G}$, that generates time reparameterizations. This last constraint consists of two terms, namely $C_\text{G}=C_\text{BI}+C_\text{inh}$. Here, $C_\text{BI}$ coincides with the Hamiltonian constraint of the LRS Bianchi I model coupled to the homogeneous massless scalar $\phi$, and $C_\text{inh}$ rules the dynamics of the inhomogeneities, coupling them with the homogeneous sector.

We then follow the hybrid approach for the quantization of this model \cite{hybrid-matter}. We adopt a Schr\"odinger representation for the homogeneous massless scalar $\phi$, a loop quantization for the Bianchi I degrees of freedom \cite{MMP,hybrid3} within the so-called improved dynamics scheme \cite{awe1}, and a Fock representation for the non-zero modes of both gravitational and matter fields \cite{Fock}. We will first deal with the representation of the Bianchi I sector.
 
\subsection{Loop quantization of the homogeneous sector}
\label{iia}

In order to adopt a loop representation of the LRS Bianchi I phase space (which is four dimensional), it is convenient to introduce four specific classical real variables to describe it, related with the non-vanishing components of the $SU(2)$ Ashtekar-Barbero connection and with the non-vanishing components of the densitized triad (see, e.g., \cite{hybrid-matter}). Following the conventions and notations of \cite{hybrid-matter,hybrid-approx}, we denote these variables as $v$, $b$, $\lambda_\theta$, and $b_\theta$. The absolute value of $v$ is proportional to the volume of the Bianchi I universe (homogeneous sector of the Gowdy model) and $\lambda_\theta$ measures the anisotropy in the $\theta$-direction. From their Poisson brackets (where $\hbar$ denotes the reduced Planck constant),
\begin{align}\label{poisson}
\{\lambda_\theta,v\}&=0,\quad \{b,v\}=\frac2{\hbar}, \quad \{b_\theta,\lambda_\theta\}=\frac2{\hbar}
\frac{\lambda_\theta}{v}, \nonumber\\
\{\lambda_\theta,b\}&=0, \quad \{b_\theta,v\}=\frac2{\hbar}, \quad \{b_\theta,b\}=\frac2{\hbar v} (b_\theta-b),
\end{align}
one observes that this set of variables is not canonical. However, it is a convenient choice to formulate the improved dynamics scheme in the loop representation \cite{awe1}. The Hilbert space resulting from this quantization is the completion of the linear span of eigenstates of the operators $\hat{v}$ and $\hat{\lambda}_\theta$ with respect to a discrete inner product. Explicitly, an orthonormal basis of the homogeneous gravitational sector is $\{|v,\lambda_\theta\rangle=|v\rangle\otimes|\lambda_\theta\rangle\}$ (with $v,\lambda_\theta\in\mathbb{R}$), where $\langle v',\lambda'_\theta|v,\lambda_\theta\rangle=\delta_{v',v}\delta_{\lambda'_\theta,\lambda_\theta}$. Let us note that, given this discrete inner product, the representation is not continuous. As a consequence, the variables $b_\theta$ and $b$ have no well-defined operator counterparts. Instead, one represents their complex exponentials, $e^{\pm ib_{\theta}}$ and $e^{\pm ib}$, that describe the holonomies of the connection:
\begin{align}\label{hol}
\widehat{e^{\pm ib_\theta}} |v,\lambda_\theta\rangle=\left|v\pm2, \lambda_\theta\pm\frac{2\lambda_\theta}{v}\right\rangle, \qquad \widehat{e^{\pm ib}} |v,\lambda_\theta\rangle=|v\pm2,\lambda_\theta\rangle.
\end{align}
Note that, with these definitions, 
\begin{align} [ \widehat{e^{\pm ib_{a}}},\widehat{v}]=i\hbar\widehat{\{e^{\pm ib_{a}},v\}},\quad\quad [ \widehat{e^{\pm ib_{a}}},\widehat{\lambda_\theta}]=i\hbar\widehat{\{e^{\pm ib_{a}},\lambda_\theta\}}, 
\end{align}
where $b_a$ denotes either $b_\theta$ or $b$. We also observe that the complex exponentials $ \widehat{e^{\pm ib_a}}$, when acting on the above basis states, produce a shift on the quantum label $v$ that is state-independent. This is the main reason motivating the above set of variables. Nevertheless, the translation on the anisotropy variable produced by $ \widehat{e^{\pm ib_\theta}}$  is not constant. This is one of the features underlying the complications when solving the dynamics of the model.

Besides, we take a standard Schr\"odinger representation for $\phi$, with Hilbert space $L^2(\mathbb{R},d\phi)$, and momentum operator $\hat{p}_\phi=-i\hbar \partial_\phi$. Then, the Bianchi I term in the Hamiltonian constraint is promoted to the following symmetric operator \cite{hybrid-matter}:
\begin{align}
\hat{C}_{\text{BI}}&= \hat{C}_{\text{FRW}}-\frac{\pi G\hbar^2}{8}(\hat{\Omega}\hat{\Theta}+\hat{\Theta}\hat{\Omega})\quad;\quad  \hat{C}_{\text{FRW}}=-\frac{3\pi G\hbar^2}{8}\hat{\Omega}^2+\frac{\hat{p}_\phi^2}{2}.
\end{align}
Here, $G$ is the Newton constant,  $\hat{\Theta}= \hat{\Theta}_\theta-\hat{\Omega}$, with
\begin{align}\label{Omegaop}
\hat{\Omega}=\sqrt{|\hat v|}\left[\widehat{{\rm sign}(v)}\widehat{\sin(b)}+\widehat{\sin(b)}\widehat{{\rm sign}(v)}\right]\sqrt{|\hat v|},
\end{align}
and $\hat{\Theta}_\theta$ is defined in a similar way, replacing $\widehat{\sin(b)}$ with $\widehat{\sin(b_\theta)}$. These operators $\hat{\Omega}$ and $\hat{\Theta}_\theta$ represent, respectively,  the classical functions $2v\sin(b)$ and $2v\sin(b_\theta)$.
The operator $\hat{C}_{\text{FRW}}$ is the Hamiltonian constraint of a flat FRW model coupled to a massless scalar $\phi$. Then, the constraint operator $\hat{C}_{\text{BI}}$ of this LRS Bianchi I sector can be seen as an FRW term plus a contribution that accounts for the anisotropies.

This $\hat{C}_{\text{BI}}$ operator has some remarkable properties thanks to the symmetric factor ordering chosen for $\hat{\Omega}$ and $\hat{\Theta}_\theta$ \cite{MMP,mmo}. It decouples the basis states $|v,\lambda_\theta\rangle$ with $v=0$ and/or $\lambda_\theta=0$ from their orthogonal complement, and it does not mix states with positive values of $v$ and/or $\lambda_\theta$ with states with negative values of those variables. This allows us to restrict our study to the subspace spanned by states $|v,\lambda_\theta\rangle$ with, e.g., $v,\lambda_\theta\in\mathbb{R}^{+}$. It is now convenient to introduce a relabeling of the basis states in this subspace as $|v,\Lambda\rangle$ ($v\in\mathbb{R}^+$, $\Lambda\in\mathbb{R}$), where $\Lambda= \ln(\lambda_\theta)$. Moreover, this subspace further splits into separable superselection sectors under the action of $\hat{C}_{\text{BI}}$. On the one hand, this action preserves all the subspaces spanned by states $|v,\Lambda\rangle$ with $v$ belonging to the semilattice of step four $\mathcal{L}_{\varepsilon}^+=\{\varepsilon+4k, k\in\mathbb{N}\}$ determined by the initial point $\varepsilon\in(0,4]$. Notice that $\varepsilon$ is the (strictly positive) minimum allowed for the Bianchi I volume $v$ in the considered sector. On the other hand, there are also superselection sectors in the anisotropy variable $\Lambda$. One can show that a state $|v,\Lambda^\star\rangle$ is related by the iterative action of the constraint only with states with $\Lambda=\Lambda^\star+\Lambda_\varepsilon$, where $\Lambda_\varepsilon$ belongs to the (countable and dense) set $\mathcal{W}_\varepsilon$ defined as \cite{hybrid3}
\begin{align}\label{W-set}
\left\{z\,\ln\left(\frac{\varepsilon-2}{\varepsilon}\right)+\sum_{m,
	n\in\mathbb{N}}{k_n^m}\ln\left(\frac{\varepsilon+2m}{\varepsilon+2n}\right);\;
k_n^m\in\mathbb{N},\; z\in\mathbb{Z}\text{ if } \varepsilon>2,\;z=0\text{ if }
\varepsilon\le2\right\}.
\end{align}
Therefore, for the homogeneous gravitational part of the system, all our future analysis will be restricted to any of the sectors spanned by basis states $|v,\Lambda\rangle$ with $v\in \mathcal L_{\varepsilon}^+$ and $\Lambda=\Lambda^\star+\Lambda_\varepsilon$, with $\Lambda_\varepsilon\in \mathcal{W}_\varepsilon$.

\subsection{Fock quantization of the inhomogeneous sector}
\label{iib}

In the hybrid strategy, a Fock quantization is adopted for the non-zero modes of the inhomogeneous fields, expressing them in terms of a suitable set of annihilation and creation-like variables, that are represented as operators acting on the corresponding Fock space. Actually, the selection of a specific quantization is possible thanks to the existence of a privileged choice of Fock representation for both the gravitational waves and the matter field, in the totally deparameterized Gowdy $T^{3}$ model. This representation is the unique one, up to unitary equivalence, that admits a unitary implementation of the dynamics and whose vacuum is invariant under rigid rotations in $\theta$, which is the gauge symmetry of the reduced system (see \cite{Fock}). This class of unitarily equivalent Fock representations requires a particular choice of configuration variables for the gravitational waves and for the matter field, that involves a homogeneous rescaling of those fields. The class contains the representation that would be most natural for massless free fields (that is, with annihilation and creation-like variables chosen as if the frequency of the modes did not include mass terms).

With this result in mind, we adopt this ``massless'' Fock quantization for the non-zero modes of the inhomogeneous fields \cite{hybrid-matter}. In this representation, we will denote the annihilation operator associated with the mode $m\in \mathbb{Z}-\{0\}$  by $\hat{a}^{(\alpha)}_m$, where $\alpha=\xi$ denotes the gravitational field and $\alpha=\varphi$ denotes the matter field, both conveniently rescaled as stated above. The Fock space admits a basis of $n$-particle states, $|\mathfrak{n}^\xi,\mathfrak{n}^\varphi\rangle$, where $\mathfrak{n}^\alpha $ denotes the infinite collection of occupation numbers $n^\alpha_{m}\in \mathbb{N}$ in each non-zero mode $m$ of the field $\alpha$. In this representation, the momentum constraint $C_\theta$ that generates rigid rotations in $\theta$ restricts the occupation numbers of the $n$-particle states so that \cite{hybrid-matter}
\begin{align}\label{mom}
\sum_{m\in\mathbb{N}^{+}}m\left(n^\xi_m+n^\varphi_m-n^\xi_{-m}-n^\varphi_{-m}\right)=0.
\end{align}

\subsection{Hamiltonian constraint of the hybrid quantum Gowdy model}
\label{iic}

The operator which represents the Hamiltonian constraint $C_\text{G}$ in this hybrid quantization is
\begin{align}\label{constra}
\hat{C}_\text{G}=\hat{C}_{\text{FRW}}-\frac{\pi G\hbar^2}{8}(\hat{\Omega}\hat{\Theta}+\hat{\Theta}\hat{\Omega})+\frac{2\pi G \hbar^{2}}{\beta} \widehat{e^{2\Lambda}}\hat{H}_0+ \frac{\pi G \hbar^2 \beta}{4}  \widehat{e^{-2\Lambda}}\hat{D}\hat{\Omega}^2\hat{D}\hat{H}_\text{I}.
\end{align}
Here,  $\beta=[G \hbar/(16\pi^{2} \gamma^{2}\Delta)]^{1/3}$ is a constant that depends on some parameters of the loop quantization ($\gamma$ is the Immirzi parameter \cite{Immirzi} and $\Delta$ is the gap in the spectrum of area eigenvalues \cite{lqg}), $\hat{H}_0$ is the free-field energy of the non-zero modes,
\begin{align}
\hat{H}_0=\sum_{\alpha\in{\xi,\varphi}}\,\sum_{m\in\mathbb{Z}-\{0\}} |m| \, \hat{a}^{(\alpha)\dagger}_m \hat{a}^{(\alpha)}_m,
\end{align}
$\hat{H}_\text{I}$ is a self-interaction term,
\begin{align}
\hat{H}_\text{I}=\sum_{\alpha\in{\xi,\varphi}}\,\sum_{m\in\mathbb{Z}-\{0\}}\frac1{2|m|} \left(2 \hat{a}^{(\alpha)\dagger}_m \hat{a}^{(\alpha)}_m+\hat{a}^{(\alpha)\dagger}_m \hat{a}^{(\alpha)\dagger}_{-m}+\hat{a}^{(\alpha)}_m \hat{a}^{(\alpha)}_{-m}\right),
\end{align}
and $\hat{D}$ represents the product of the volume by its inverse, which is regularized in LQC \cite{lqc}. This product differs from the identity only in the region of small volumes:
\begin{align}\label{Dop}
\hat{D}|v\rangle=D(v)|v\rangle,\qquad D(v)= v\left(\sqrt{v+1}-\sqrt{|v-1|}\right)^2.
\end{align}
The last two terms in \eqref{constra} form the operator $\hat{C}_\text{inh}$ that represents the inhomogeneous term $C_\text{inh}$ of the constraint. Given its action on the $|v,\Lambda\rangle$-part of the states, the Hamiltonian constraint $\hat{C}_\text{G}$ preserves the superselection sectors introduced in Subsec. \ref{iia}. Actually, it is densely defined on the kinematical Hilbert space defined by the tensor product of $L^2(\mathbb{R},d\phi)$ with the completion of the linear span of $|v,\Lambda,\mathfrak{n}^\xi,\mathfrak{n}^\varphi\rangle$ with respect to the inner product 
\begin{align}
\langle v',\Lambda',\mathfrak{n'}^\xi,\mathfrak{n'}^\varphi|v,\Lambda,\mathfrak{n}^\xi,\mathfrak{n}^\varphi\rangle=\delta_{v',v}\delta_{\Lambda',\Lambda}\delta_{\mathfrak{n'}^\xi,\mathfrak{n}^\xi}\delta_{\mathfrak{n'}^\varphi,\mathfrak{n}^\varphi}, 
\end{align} 
where we recall that $v\in \mathcal L_{\varepsilon}^+$ and $\Lambda=\Lambda^\star+\Lambda_\varepsilon$, with $\Lambda_\varepsilon\in \mathcal{W}_\varepsilon$.

\section{Approximating the Hamiltonian constraint}
\label{sec:app}

The action of the Hamiltonian constraint $\hat{C}_\text{G}$ on the kinematical Hilbert space is quite complicated, a fact that makes very difficult (if not impossible) to analytically find the physical states of the system, those that solve the equation $\hat{C}_{\text{G}} |\Psi\rangle=0$.\footnote{Here $|\Psi \rangle$ generally stands for a generalized state. Alternatively, we can understand our constraint as an equation of the form $(\Psi|\hat{C}_{\text{G}}^{\dagger} $ on generalized states on the dual of the domain of $\hat{C}_{\text{G}}$.} This is due to the presence of two  terms: The anisotropy term containing $\hat{\Omega}\hat{\Theta}+\hat{\Theta}\hat{\Omega}$ and the interaction term containing $\widehat{e^{-2\Lambda}}\hat{D}\hat{\Omega}^2\hat{D}\hat{H}_\text{I}$. On the one hand, as far as the homogeneous sector is concerned, neither the anisotropy operator $\hat{\Omega}\hat{\Theta}+\hat{\Theta}\hat{\Omega}$ nor the operator $\hat{D}$ commute with the FRW operator $\hat{\Omega}^2$. Therefore they cannot be diagonalized simultaneously. This fact, together with the $v$-dependent translations that the anisotropy operator produces in $\Lambda$ [see \eqref{hol}], makes the resolution of the constraint in the homogeneous variables $v$ and $\Lambda$ extremely hard. On the other hand, concerning the inhomogeneous sector, the interaction operator $\hat{H}_\text{I}$, which is densely defined on the linear span of the $n$-particle states, does not act diagonally on them (in contrast with $\hat{H}_0$). Specifically, it creates and annihilates a pair of particles in every mode, notably complicating the resolution of the constraint for the inhomogeneous degrees of freedom.

Owing to these obstacles, in what follows we will carry out approximations that allow us to disregard those problematic terms when acting on certain families of states. This will lead to a much simpler constraint operator, possessing some solutions that can be regarded as approximate solutions to the full Gowdy model, and which will be studied in Sec. \ref{sec:sol}.

Let us start by reviewing the spectral properties of the operator $\hat{\Omega}^2$. It is essentially self-adjoint, with an absolutely continuous, non-degenerate, and positive spectrum \cite{mmo}. In each of the superselection sectors, with support of $v$ in a semilattice $\mathcal{L}^+_\varepsilon$, the delta-normalized eigenstates of $\hat{\Omega}^2$ with eigenvalue $\rho^2\in \mathbb{R}^+$,
\begin{align}
|e^\varepsilon_\rho\rangle=\sum_{v\in\mathcal{L}^+_\varepsilon}e^\varepsilon_\rho(v)|v\rangle,
\end{align} 
provide a resolution of the identity. The eigenfunctions can be chosen to be real and present a feature that will prove essential for our future approximations, namely, that when $\rho \gg 10$, $e^\varepsilon_\rho(v)$ is exponentially suppressed for $v\lesssim \rho/2$. On the other hand, for $v\gg \rho/2$, these eigenfunctions are oscillatory (see, e.g., \cite{hybrid-approx}). The exponential suppression of the region  $v\lesssim \rho/2$ is  a characteristic of the quantum geometry effects in the context of LQC, and it is at the root of the occurrence of a quantum bounce in the loop quantization \cite{mmo}. In the proximity of this bounce, the LQC phenomena alter significantly the gravitational behavior with respect to the predictions of General Relativity, invalidating the expectations based on this latter theory \cite{effective}. Actually, this bounce resolves the cosmological singularities. Moreover, it persists in the presence of anisotropies \cite{param} and of inhomogeneities \cite{tarrio}. 

Let us now  consider states whose homogeneous gravitational contribution has the form 
\begin{align} |\mathcal{G}\rangle=\sum_{\Lambda_\varepsilon\in\mathcal{W}_\varepsilon}\int_{\text{Spc}(\omega)} \text{d}\omega \, g(\omega,\Lambda^\star+\Lambda_\varepsilon)|\omega,\Lambda^\star+\Lambda_\varepsilon\rangle, 
\end{align} 
with
\begin{align}\label{ani-omega}
g(\omega,\Lambda)=N(\omega) f(\omega,\Lambda),\qquad f(\omega,\Lambda)= e^{-\frac{\sigma^{2}_{s}}{2q_{\epsilon}^{2}}[\Lambda-\bar{\Lambda}(\omega)]^{2}}.
\end{align}
Here,
\begin{align}\label{scale}
q_\varepsilon=\ln\left(1+\frac{2}{v_m}\right),
\end{align}
with $v_m\in\mathcal L^+_\varepsilon$ a certain value of the volume such that $v_m\gg 10$. Besides, $\sigma_s$ is a free parameter characterizing the width of the Gaussian $f(\omega,\Lambda)$ (together with $q_\varepsilon$), and $\text{Spc}(\omega)$ denotes the spectrum of some operator $\hat{\omega}$ defined on the homogeneous and isotropic geometry part of the kinematical Hilbert space (that is, the space spanned by the states $| v\rangle$). We will assume $\hat{\omega}$ to be (essentially) self-adjoint. Let us notice that its spectrum might be continuous, discrete, or even a mixture of both types. Nevertheless, we will formally denote the spectral resolution of the identity provided by $\hat{\omega}$ as $\mathbb{I}=\int_{\text{Spc}(\omega)} \text{d}\omega |\omega\rangle \langle \omega|$. States $| \mathcal{G}\rangle$ of this kind were studied in \cite{hybrid-approx}, though only in two particular cases: First when $\hat{\omega}$ is a constant operator, and then when $\hat{\omega}=\hat{v}$. 
Moreover, keeping in mind the key properties of the states in \cite{hybrid-approx} needed for the different approximations introduced in those references, we will for the moment assume that the profile in the $\rho$-representation,
\begin{align} 
 g(\rho,\Lambda)=\int_{\text{Spc}(\omega)} \text{d}\omega \, g(\omega,\Lambda) e^\varepsilon_{\omega}(\rho), 
\end{align} 
is highly suppressed in the region with $\rho \lesssim \rho_m = 2v_m$. Here we are denoting by $e^\varepsilon_{\omega}(\rho)$ the wave function of the state $|\omega\rangle$ in the $\rho$-representation,
$e^\varepsilon_{\omega}(\rho)=\langle e^\varepsilon_\rho | \omega \rangle.$
Owing to the exponential suppression of $e^\varepsilon_\rho(v)$ in the region $v\lesssim \rho/2$ discussed in the previous paragraph,  a profile $g(\rho,\Lambda)$ highly suppressed for $\rho \lesssim \rho_m$ in turn implies that the corresponding profile in the $v$-representation, $g(v,\Lambda)=\int_{0}^\infty \text{d}\rho \, g(\rho,\Lambda) e^{\varepsilon}_{\rho}(v)$, is highly suppressed for $v \lesssim v_m$.

In the following, we will analyze under which conditions the approximations of \cite{hybrid-approx} extend to these generalized states with homogeneous profiles characterized by \eqref{ani-omega}. Later on, in Sec. \ref{sec:sol}, we will explain how to construct approximate solutions with the desired properties.

\subsection{Approximating the anisotropy term}
\label{iiia}

In order to deal with the anisotropy operator $\hat{\Omega}\hat{\Theta}+\hat{\Theta}\hat{\Omega}$, it is essential to notice that, when acting on the considered states $| \mathcal{G}\rangle$, and owing to the suppression in the region $v \lesssim v_m$ that we are assuming, the only contributing $v$-dependent shifts on the anisotropy variable $\Lambda$ that its action produces are not (significantly) bigger than the scale $q_\varepsilon$ \cite{hybrid-approx}. Recalling that the considered profiles $g(\omega,\Lambda)$ are Gaussian-like in the anisotropy variable with width given by $q_\varepsilon/\sigma_{s}$, if that scale is much smaller than their width, namely if $\sigma_{s}\ll 1$, then they can be extended (from the anisotropy superselection sector) to a smooth function in $\Lambda$ so that, for contributing shifts $\Lambda_{0}\leq q_{\varepsilon}$,
\begin{align}\label{tayl}
g(\omega,\Lambda+\Lambda_{0})\simeq g(\omega,\Lambda) +\Lambda_{0}\partial_{\Lambda} g(\omega,\Lambda) ,
\end{align}
for all $\omega$ in the support of $g(\omega,\Lambda)$. If one considers the $v$-representation of $| \mathcal{G}\rangle$, this in turn implies that $g(v,\Lambda+\Lambda_{0})\simeq g(v,\Lambda) +\Lambda_{0}\partial_{\Lambda} g(v,\Lambda)$ for all $v$.

For such states,  the operator $\hat{\Omega}\hat{\Theta}+\hat{\Theta}\hat{\Omega}$ approximately factorizes as $-2\hat{\tilde\Omega}\hat{\Theta}'$. Here, $\hat{\tilde\Omega}$ is defined like the geometry operator $\hat{\Omega}$ in \eqref{Omegaop}, except for the substitution of the conjugate pair of variables $(v,b)$ by $(v/2,2b)$.
On the other hand, $\hat{\Theta}'$ is the discretization of the first derivative $-4i\partial_\Lambda$ at the scale $q_\varepsilon$:
\begin{align}\label{zetaprim}
\hat{\Theta}'|\Lambda\rangle= i\frac2{q_\varepsilon}\left(|\Lambda+q_\varepsilon\rangle-|\Lambda-q_\varepsilon\rangle\right).
\end{align}
The proof of this approximation follows exactly the same steps  as in \cite{hybrid-approx}. For further details, we refer the reader to that work.

Note that $q_\varepsilon\in\mathcal{W}_\varepsilon$ so that $\hat{\tilde\Omega}\hat{\Theta}'$ preserves the  superselection sectors where  $\hat{\Omega}\hat{\Theta}+\hat{\Theta}\hat{\Omega}$ is defined. Furthermore, the action of $\hat{\Theta}'$ leaves invariant the lattices of constant step $q_\varepsilon$ of the form $\mathcal{L}^{q_\varepsilon}_{\Lambda^\prime} =\{ \Lambda^\prime+nq_\varepsilon;\, n\in\mathbb{Z}\}$ contained in those sectors, with  
$\Lambda^\prime-\Lambda^\star\in \mathcal{W}_\varepsilon$. Using this result, we can further restrict the study of the anisotropy variable just to the subspace spanned by states $|\Lambda\rangle$ with support in any of those lattices.

We will now show that the approximate anisotropy operator given by $-2\hat{\tilde\Omega}\hat{\Theta}'$ can be disregarded when acting on the considered states $| \mathcal{G}\rangle$ in comparison with $\hat{\Omega}^2$. First of all, it is easy to check that our definition of the operator $\hat{\tilde\Omega}$ is equivalent to the following action on the basis states:
\begin{align}
\hat{\tilde\Omega}|v\rangle= i \left[\tilde{y}_-(v)|v-4\rangle - \tilde{y}_+(v)|v+4\rangle\right],
\end{align}
with
\begin{align}
 \tilde{y}_\pm(v)=\frac{1+\text{sign}(v\pm4)}{4}\sqrt{v(v\pm 4)}.
\end{align}
Employing this, we get that the approximate action of the negative of $\hat{\Omega}\hat{\Theta}+\hat{\Theta}\hat{\Omega}$ on our states is given by
\begin{align}\label{form315}
\langle v,\Lambda|2 \hat{\tilde\Omega}\hat{\Theta}'| \mathcal{G}\rangle=&\frac{4}{q_\varepsilon}\big\{\tilde{y}_+(v)\left[g(v+4,\Lambda+q_\varepsilon)-g(v+4,\Lambda-q_\varepsilon)
\right]\nonumber\\-&    \tilde{y}_-(v)\left[g(v-4,\Lambda+q_\varepsilon)-g(v-4,\Lambda-q_\varepsilon)\right]\big\}.
\end{align}
Now, given the form \eqref{ani-omega} of the profiles, a straightforward calculation leads to
\begin{multline}
\frac{2}{q_\varepsilon}\big[g(v\pm 4,\Lambda+q_\varepsilon)-g(v\pm 4,\Lambda-q_\varepsilon)\big]
\\
= -\frac{4}{q_\varepsilon} e^{-\frac{\sigma_s^2}{2} }\int_{\text{Spc}(\omega)} \text{d}\omega \, g(\omega,\Lambda) e^\varepsilon_{\omega}(v\pm 4) \sinh\left(\frac{\sigma_s^2}{q_\varepsilon}[\Lambda-\bar\Lambda(\omega)]\right).
\end{multline}
Here, $e^\varepsilon_{\omega}(v)=\langle v | \omega \rangle$ denotes the wave function of the state $|\omega\rangle$ in the $v$-representation. The function $g(\omega,\Lambda)$ only contributes when ${\sigma_s}[\Lambda-\bar\Lambda(\omega)]
/{q_\varepsilon}$ is $\mathcal{O}(1)$. Therefore, we get
\begin{align}
\frac{2}{q_\varepsilon}\big[g(v\pm 4,\Lambda+q_\varepsilon)-g(v\pm 4,\Lambda-q_\varepsilon)\big]\simeq -\frac{4\sinh\sigma_s}{q_\varepsilon} e^{-\frac{\sigma_s^2}{2} } g(v\pm 4,\Lambda) \times \mathcal{O}(1).
\end{align}
Hence, recalling \eqref{form315}, we have just obtained that the action of the Bianchi I anisotropy operator on these states can be approximated as follows, when one stays in the sector $v \gtrsim v_m \gg 10$ under consideration:
\begin{align}\label{aproequani}
\langle v,\Lambda| \hat{\Omega}\hat{\Theta}+\hat{\Theta}\hat{\Omega} | \mathcal{G}\rangle \simeq 8\frac{\sinh\sigma_s}{q_\varepsilon} e^{-\frac{\sigma_s^2}{2} } \langle v,\Lambda| \hat{\tilde\Omega} | \mathcal{G}\rangle,
\end{align}
up to a factor $\mathcal{O}(1)$ on the right-hand side.
Taking into account that the action of $\hat{\tilde\Omega}$ on  $| \mathcal{G}\rangle$ is of the same order as that of $\hat{\Omega}$, since these operators are completely analogous except for the magnitude of the shifts that they produce in $v$, we can expect the right-hand side of \eqref{aproequani} to be negligible compared to $\langle v,\Lambda|\hat{\Omega}^2| \mathcal{G}\rangle$ if $8 |\sinh\sigma_s| \exp{(-\sigma_s^2/2)} \ll q_\varepsilon v_m$. Here, we have used the fact that, in the region with $v \gtrsim v_m \gg 10$, 
\begin{align}\label{appOmega2act}
\hat{\Omega}^2| v\rangle=-{y}_{++}(v)| v+4\rangle +2v^2 | v\rangle -{y}_{--}(v)| v-4\rangle,
\end{align}
with
\begin{align}\label{appdefini2}
y_{++}(v)=y_+(v) y_+(v+2),\qquad y_{--}(v)=y_{++}(v-4),
\end{align}
\begin{align}\label{appdef2}
y_{+}(v)=\frac{1+\text{sign}(v+2)}{2}\sqrt{v(v+2)},
\end{align}
and so the coefficients of $\langle v,\Lambda|\hat{\Omega}^2| \mathcal{G}\rangle$ are $\mathcal O(v^2)$. Equivalently, since for $v_m \gg 10$, $q_\varepsilon \simeq 2/v_m$, we can rewrite the above condition as
\begin{align}\label{anicondsigma}
4\left|\sinh\sigma_s\right| e^{-\frac{\sigma_s^2}{2} } \ll 1.
\end{align}
Recalling that we have already required that $\sigma_s \ll 1$ in order to treat $g(v, \Lambda)$ as smooth in $\Lambda$ and approximate it by its Taylor expansion truncated at the first derivative, the new condition \eqref{anicondsigma} turns out to be trivially satisfied in the considered case.

This concludes the proof that the action of the anisotropy term on states with profile of the form \eqref{ani-omega} can be disregarded in the Hamiltonian constraint provided that the condition $\sigma_s \ll 1$ is satisfied and that $g(\rho,\Lambda)$ is highly suppressed for $\rho \lesssim \rho_m$. This was actually expected, since the Gaussian-like profiles that we are considering for the anisotropies are reasonably centered on trajectories with vanishing momenta $\hat{\Theta}'$.

\subsection{Approximating the interaction term}
\label{iiib}

Let us analyze now the interaction term $\widehat{e^{-2\Lambda}}\hat{D}\hat{\Omega}^2\hat{D}\hat{H}_\text{I}$. Before discussing whether this operator can be neglected in the Hamiltonian constraint when considering states $| \mathcal{G}\rangle$ with profiles of the form \eqref{ani-omega}, it is worth noticing that, owing to their suppression in the sector $v \lesssim v_m$ with $v_m \gg 10$, sector on which $\hat{D}$ just acts like the identity according to \eqref{Dop}, we can make the approximation:
\begin{align}\label{approxD}
\widehat{e^{-2\Lambda}}\hat{D}\hat{\Omega}^2\hat{D}\hat{H}_\text{I}| \mathcal{G}\rangle\simeq  \widehat{e^{-2\Lambda}}\hat{\Omega}^2\hat{H}_\text{I}| \mathcal{G}\rangle.
\end{align}
Thus, provided that the content of inhomogeneities of our states is reasonable, we can disregard this interaction term if the action of $\widehat{e^{-2\Lambda}}\hat\Omega^2$ on them is negligible compared to the rest of contributions in the Hamiltonian constraint. This action is given by
\begin{align}
\widehat{e^{-2\Lambda}}\hat{\Omega}^2| \mathcal{G}\rangle
=\sum_{\Lambda\in\mathcal{L}^{q_\varepsilon}_{\Lambda^\prime}}\int_{\text{Spc}(\omega)} \text{d}\omega \, N(\omega) e^{-\frac{\sigma_{s}^{2}}{2q_{\varepsilon}^{2}}\left[\Lambda-\bar{\Lambda}(\omega)+\frac{2q_{\varepsilon}^{2}}{\sigma_{s}^{2}}\right]^{2}}
e^{-2\bar{\Lambda}(\omega)
+\frac{2q_{\varepsilon}^{2}}{\sigma_{s}^{2}}}\hat{\Omega}^2| \omega\rangle.
\end{align}
Note that the first exponential of the right-hand side is bounded from above by the unit. Therefore we see that, if we choose the peak of the Gaussian-like profiles such that $\bar\Lambda(\omega)$ is much bigger than both  $1$ and $q_\varepsilon^2/\sigma_s^2$ for all $\omega$ in the support of $N(\omega)$, then this contribution will be negligible compared to $\hat{\Omega}^2| \mathcal{G}\rangle$. In conclusion, we can disregard the interaction term in the Hamiltonian constraint of the Gowdy model for states with the considered profiles if it is satisfied that $\bar\Lambda(\omega)\gg \text{max}\left(1,q_\varepsilon^2/\sigma_s^2\right)$ for all values of $\omega$ in the support of the state (where $\text{max} $ stands for the maximum).

\subsection{Approximating the free term}
\label{iiic}

There is one more approximation that can be made in the Hamiltonian constraint $\hat{C}_\text{G}$ on states $| \mathcal{G}\rangle$ with the profile \eqref{ani-omega}. This is an approximation for the term containing the free contribution of the inhomogeneities, namely (up to constants) $\widehat{e^{2\Lambda}}\hat{H}_0$. The idea is to restrict the Gaussian-like profiles in $g(\omega,\Lambda)$ to be sharply peaked at $\bar\Lambda(\omega)$, that is, to require its width $q_\varepsilon/\sigma_{s}$ to be much smaller than the unity. If this is so, then
\begin{align}\label{mimick}
\widehat{e^{2\Lambda}}| \mathcal{G}\rangle \simeq e^{2\bar\Lambda(\hat{\omega})}| \mathcal{G}\rangle.
\end{align}
Indeed, the support of a Gaussian is approximately its width, and in our case this support corresponds to $\Lambda$ such that $| \Lambda-\bar\Lambda(\omega) |\leq q_\varepsilon/\sigma_{s}$. Therefore, the non-negligible contributions of the action of $\widehat{e^{2\Lambda}}$ on these states will take the values
\begin{align}
e^{2\Lambda}=e^{2\bar\Lambda(\omega)+ 2\alpha\frac{q_\varepsilon}{\sigma_s}}, \qquad \alpha\in [-1,1].
\end{align}
Thus, if $q_\varepsilon/\sigma_{s} \ll 1$, this value will be essentially equal to that taken by the operator $e^{2\bar\Lambda(\hat{\omega})}$ on our state (term by term in its spectral decomposition). Then, this approximation will be consistent with disregarding the anisotropy term if and only if the parameters that characterize the width of the Gaussian-like profile are such that $q_\varepsilon \ll \sigma_s \ll 1$. Let us conclude noticing that, in this way, by considering states with profiles for the anisotropy sharply peaked at a function of some operator of the homogeneous and isotropic geometry, one ends up mimicking a contribution in the approximated Hamiltonian constraint that is given precisely by that very same operator, in the sense of  \eqref{mimick}.

\section{Approximate solutions to the Gowdy constraint: modeling modified FRW cosmologies}
\label{sec:sol}

In the previous section, we have shown how the full Gowdy Hamiltonian constraint $\hat{C}_\text{G}$ can be approximated by the operator
\begin{align}\label{const-app-prime}
\hat{C}^{\prime}_{\text{app}}=\hat{C}_{\text{FRW}}+\frac{2\pi G \hbar^{2}}{\beta} e^{2\bar\Lambda(\hat{\omega})}\hat{H}_0=-\frac{3\pi G\hbar^2}{8}\hat{\Omega}^2+\frac{\hat{p}_\phi^2}{2}+\frac{2\pi G \hbar^{2}}{\beta} e^{2\bar\Lambda(\hat{\omega})}\hat{H}_0
\end{align}
when acting on quantum states whose homogeneous gravitational part is given by 
\begin{align}
|\mathcal{G}\rangle=\sum_{\Lambda\in\mathcal{L}^{q_\varepsilon}_{\Lambda^\prime}}\int_{\text{Spc}(\omega)} \text{d}\omega \,  g(\omega,\Lambda)|\omega,\Lambda\rangle , 
\end{align} 
with $g(\omega,\Lambda)$ defined in \eqref{ani-omega}, provided that the following conditions are satisfied:
\begin{itemize}
\item[i)] In the $\rho$-representation, $g(\rho,\Lambda)$ has to be highly suppressed for $\rho \lesssim \rho_m$, with $\rho_m\gg 10$. This in particular implies that $g(v,\Lambda)$ is exponentially suppresed for $v \lesssim v_m  = \rho_m/2$.
\item[ii)] $\bar\Lambda(\omega)\gg 1$ for all $\omega$ in the support of $N(\omega)$.
\item[iii)] $q_\varepsilon \ll \sigma_s \ll 1$.
\end{itemize}
In what follows, we will consider solutions of this approximate constraint that are in turn approximate solutions of the full Gowdy model. Later on, a possible physical interpretation of those solutions will be given.

\subsection{Construction of solutions}
\label{iva}

Let us search for solutions of the approximate constraint $\hat{C}^{\prime}_{\text{app}}|\Psi \rangle=0$ by considering states
\begin{align}\label{gen-states}
|\Psi\rangle=\int_{-\infty}^{\infty} \text{d}{p_\phi}\int_{\text{Spc}(\omega)} \text{d}\omega \sum_{\Lambda\in\mathcal{L}^{q_\varepsilon}_{\Lambda^\prime}}
\sum_{\mathfrak{n}^{\xi},\mathfrak{n}^{\varphi}}\Psi(p_\phi,\omega,\Lambda,\mathfrak{n}^{\xi},\mathfrak{n}^{\varphi})
| p_\phi,\omega,\Lambda,\mathfrak{n}^{\xi},\mathfrak{n}^{\phi}\rangle
\end{align} with wave function of the form
\begin{align}\label{wave function}
\Psi(p_\phi,\omega,\Lambda,\mathfrak{n}^{\xi},\mathfrak{n}^{\varphi})&= f(\omega,\Lambda)N(\omega,p_\phi,\mathfrak{n}^{\xi},\mathfrak{n}^{\varphi}),
\end{align}
where $f(\omega,\Lambda)$ is given in \eqref{ani-omega} and satisfies by construction our condition iii) above. The sets of occupation numbers  $\mathfrak{n}^{\xi}$  and $\mathfrak{n}^{\varphi}$ determine the eigenvalue of $\hat{H}_0$ (which acts diagonally on the $n$-particles states):
\begin{align}
H_{0}\equiv H_{0}(\mathfrak{n}^{\xi},\mathfrak{n}^{\varphi})=\sum_{m\in\mathbb{Z}-\{0\}}|m|(n^\xi_m+n^\varphi_m).
\end{align} 
Here, we have generically included in $N(\omega)$ the dependence of the wave function on $p_\phi$ and on these occupation numbers.  This function $N(\omega,p_\phi,\mathfrak{n}^{\xi},\mathfrak{n}^{\varphi})$ should be chosen in such a way that the content of inhomogeneities is small, so that the approximation of disregarding $\hat{C}_\text{I}$ holds, and such that the momentum constraint \eqref{mom} is satisfied.

Since $\hat{p}_\phi$ and $\hat{H}_0$ are Dirac observables of this approximate constraint, and therefore $p_\phi$ and $H_{0}$ are constants of motion, solving the approximate constraint $\hat{C}^{\prime}_{\text{app}}|\Psi\rangle=0$ on the above states is equivalent to solve  $\hat{C}_{\text{app}}|\Psi\rangle =0$ in each eigenspace of the two considered Dirac observables, with $\hat{C}_{\text{app}}$ being the operator
\begin{align}\label{const-app}
\hat{C}_{\text{app}}=-\frac{3\pi G\hbar^2}{8}\hat{\Omega}^2+\frac{p_\phi^2}{2}+\frac{2\pi G \hbar^{2}}{\beta} e^{2\bar\Lambda(\hat{\omega})}H_0.
\end{align}
Note that this constraint operator only acts on the homogeneous and isotropic part of the Hilbert space, namely that of the FRW model. However, let us emphasize the fact that, to each state $\Psi$, there corresponds a (collection of operators) $\hat{C}_{\text{app}}$, as this operator depends (besides as on $p_\phi$ and $H_0$) on $e^{2\bar\Lambda(\omega)}$.

Keeping in mind that we are interested in states that are in turn approximate solutions of the full Gowdy model, the wave function \eqref{wave function} must be such that these states satisfy conditions i) and ii) above as well. In order to fulfill these conditions, and motivated by the strategy employed in \cite{hybrid-approx}, we will restrict all considerations to states with a peak of the Gaussian $\bar\Lambda(\omega)$ such that the resulting operator $\bar\Lambda(\hat{\omega})$ in \eqref{const-app} is defined in the following way:
\begin{align}\label{lamb}
\bar\Lambda(\hat{\omega}) =
\begin{cases}
\bar\Lambda_0, & \text{if }v< v_0, \\
\bar\Lambda(\hat{\omega}_0), & \text{if }v\geq v_0,
\end{cases}
\end{align}
for certain $v_0\geq v_m$. Here, $\bar\Lambda_0$ is a constant and $\hat{\omega}_0$ is the restriction (via projection) of $\hat{\omega}$ to the sector of the homogeneous and isotropic part of the Hilbert space with $v\geq v_0$. Explicitly, $\hat{\omega}_0=\hat{P}\hat{\omega}\hat{P}$, where $\hat{P}$ is the projector on the linear span of $| v\rangle$ with $v\in \mathcal{L}^{p}_{\varepsilon} = \mathcal{L}^+_{\varepsilon}\cap\{v\geq v_0\}$. 
Note that $v_0$ does not necessarily belong to the semilattice $\mathcal{L}^+_{\varepsilon}$ considered. For future reference, let us denote by $v_1$ the lowest end of $\mathcal{L}^{p}_{\varepsilon}$, so that $v_0\leq v_1<v_0+4$.
Furthermore, we will focus on operators $\hat\omega$ such that
\begin{align}\label{Op}
e^{2\bar\Lambda(\hat{\omega}_0)}=e^{2h(\hat{v})}+\hat{O}_{p},
\end{align}
where the function $h(v)$ varies sufficiently smoothly so as to allow us to make the approximation $h(v\pm 4)\simeq h(v)$ in the region $v\geq v_0$. Besides, we assume $\hat{O}_{p}$ to be a positive (and thus self-adjoint) operator defined on the linear span of $| v\rangle$ with $v\in\mathcal{L}^{p}_{\varepsilon}$, with a quasi-local action on this basis of the generic form
\begin{align}\label{Opact}
\hat{O}_{p} | v\rangle=\sum_{k=0}^{K}\left[f_{k}^{+}(v) | v+4k\rangle + f_{k}^{-}(v) | v-4k\rangle\right], \qquad K<\infty.
\end{align}
Here, $f_{0}^{+}=f_{0}^{-}\equiv f_{0}$ and $f_{k}^{-}(v)=0$ if $v-4k<v_{0}$, so that $\hat{O}_{p}$ is indeed defined on the sector $v\geq v_0$. The case with $\hat{O}_{p}=0$ was the one studied in \cite{hybrid-approx}. For conciseness, we now assume that $f_K^+(v)$ is (strictly) positive for $v\geq v_0$. As shown in Appendix \ref{appendixA}, states $\Psi$ that solve the constraint equation $ \hat{C}_{\text{app}}|\Psi\rangle=0$ can be determined from the equation
\begin{align}\label{const-N}
\sum_{v\in\mathcal{L}^+_{\varepsilon}} N(v,p_\phi,\mathfrak{n}^{\xi},\mathfrak{n}^{\varphi})\hat{C}_{\text{app}} | v\rangle=0,
\end{align}
with
\begin{align} N(v,p_\phi,\mathfrak{n}^{\xi},\mathfrak{n}^{\varphi})= \int_{\text{Spc}(\omega)} \text{d}\omega \, N(\omega,p_\phi,\mathfrak{n}^{\xi},\mathfrak{n}^{\varphi}) e^\varepsilon_{\omega}(v).
\end{align} 

We will now analyze how we can construct solutions to \eqref{const-N} with the desired properties. First of all, condition ii) is automatically satisfied for these states if $\bar\Lambda_0 \gg 1$ and $h(v) \gg 1$ for all $v\in\mathcal{L}^{p}_{\varepsilon}$. Concerning condition i), for $v< v_{0}$ the operator $\bar\Lambda(\hat{\omega})$ is simply a constant and the constraint equation reduces to an eigenvalue equation for the FRW operator $\hat{\Omega}^2$, given by \cite{hybrid-approx}
\begin{align}\label{rhosqu}
\rho^2(p_\phi,\bar\Lambda_0,H_0)&=\frac{4}{3\pi G\hbar^2}p_\phi^{2}+\frac{16}{3\beta} e^{2\bar\Lambda_0}H_{0}(\mathfrak{n}^{\xi},\mathfrak{n}^{\varphi}).
\end{align}
Therefore, for all $v \leq v_1$ solutions are of the form 
\begin{align}\label{oldsoluA}
N(v,p_\phi,\mathfrak{n}^{\xi},\mathfrak{n}^{\varphi})= e^{\varepsilon}_{\rho(p_\phi,\bar\Lambda_0,H_0)}(v)
\psi(p_\phi,\mathfrak{n}^{\xi},\mathfrak{n}^{\varphi}),
\end{align} 
where we recall that $e^{\varepsilon}_{\rho}$ is the wave function in the $v$-representation of the eigenfunction of  $\hat{\Omega}^2$ with eigenvalue $\rho^2$. Hence, condition i) for these states is satisfied, for instance, if one restricts the function $\psi(p_\phi,\mathfrak{n}^{\xi},\mathfrak{n}^{\varphi})$ to have support on the region with $p_\phi> p_{\phi}^m$, where we choose $ p_{\phi}^m \gg\sqrt{75\pi G}\hbar$. That this is true follows from the positivity of the last term in \eqref{rhosqu}. Indeed, in that case, solutions will have significant contributions only for
\begin{align}\label{rhomdef}
\rho\geq  \frac{2}{\sqrt{3\pi G}\hbar}p_\phi > \frac{2}{\sqrt{3\pi G}\hbar}p_\phi^m =\rho_m \gg 10.
\end{align}
The identity in this formula is just a definition of the scale $\rho_m$ used in our construction of states. Let us note that this scale (and therefore the corresponding value of $v_{m}$) has been defined in an intrinsic way, in terms of the conserved momentum of the homogeneous scalar field. 

In the region with $v\in\mathcal{L}^{p}_{\varepsilon}$, equation \eqref{const-N} leads to the difference equation [dropping the dependence of $N(v,p_\phi,\mathfrak{n}^{\xi},\mathfrak{n}^{\varphi})$ on $p_\phi$ and the occupation numbers in order to simplify the notation]
\begin{align}\label{diff}
\left[\frac{4p_\phi^2}{3\pi G\hbar^2}+\frac{16}{3\beta} e^{2\tilde\Lambda(v)}H_0-2v^2\right]N(v)+y_{++}(v)N(v+4)+y_{--}(v)N(v-4)\nonumber \\+\frac{16}{3\beta}H_0\sum_{k=1}^{K}\left[f_{k}^{+}(v-4k) N(v-4k)\chi_{\mathcal{L}^{p}_{\varepsilon}}(v-4k) + f_{k}^{-}(v+4k) N(v+4k)\right]=0.
\end{align}
 Here,
\begin{align}
\chi_{\mathcal{L}^{p}_{\varepsilon}}(v)=
\begin{cases}
1, & \text{if }v\in\mathcal{L}^{p}_{\varepsilon}, \\
0, & \text{if }v\notin\mathcal{L}^{p}_{\varepsilon},
\end{cases}
\end{align}
and we have defined the function
\begin{align}
\tilde\Lambda(v)=\frac{1}{2}\ln\left[e^{2h(v)}+2f_{0}(v)\right].
\end{align}
The difference equation \eqref{diff} involves coefficients $N(v)$ evaluated on $K$ points above the considered one, $v$, in the semilattice. In particular, this happens at the matching point with the solution of constant $\bar{\Lambda}_0$, i.e., when $v$ equals the lower end $v_1$ of $\mathcal{L}^{p}_{\varepsilon}$. Owing to this fact and that, at this matching point, we know data only for values of $v$ smaller than or equal to it, it is not difficult to realize that, in order to be able to find an approximate solution to our equation without ambiguities, we must impose certain requirements on the operator $\hat{O}_{p}$. For instance, to arrive at acceptably smooth solutions, we can require that the functions $f_{k}^{\pm}(v)$ with $k\neq 0$ be negligible in a neighbourhood above $v_{0}$, namely, at least at all points in the interval $\mathcal{I}=[v_{0},v_{0}+8K-4)$. Indeed, if this is so and after disregarding the contribution of those functions, then \eqref{diff}  approximately gives us in a deterministic way the $K-1$ values of the function $N(v)$ from $v=v_{1}+4$ up to $v=v_{1}+4K-4$, when supplemented with the input data 
\begin{align}
 e^{\varepsilon}_{\rho(p_\phi,\bar\Lambda_0,
H_0)}(v_1) \qquad {\rm and} \qquad e^{\varepsilon}_{\rho(p_\phi,\bar\Lambda_0,
H_0)}(v_1-4), 
\end{align}
coming from the imposition of the constraint in the region $v<v_{0}\leq v_1$. We note, in particular, that to fix $N(v_1+4K-4)$ in this manner we need to ignore the functions $f_{k}^{\pm}(v)$ at least up to $v_1+8K-8$, for any possible value of $v_1\in [v_0, v_0+4)$, as we have assumed above that it is indeed the case. The $K$ coefficients computed in $\mathcal{L}^{p}_{\varepsilon}$, from $N(v_1)$ to $N(v_1+4K-4)$, can then serve as initial data in order to uniquely fix the rest of coefficients. The procedure to do so is to consider again the full constraint equation \eqref{diff} evaluated at points $v\geq v_1$ without neglecting the functions $f_{k}^{\pm}(v)$. Introducing the value of those $K$ coefficients, the constraint for $v=v_1$ completely determines the next coefficient $N(v_1+4K)$, and so on and so forth for $v\geq v_1+4$. With this method, the whole solution can be constructed, at least approximately.

Let us notice that, if the neighbourhood $\mathcal{I}$ contains $n>2K-1$ points of $\mathcal{L}^{p}_{\varepsilon}$, then the number of coefficients that can be obtained by ignoring the functions $f_{k}^{\pm}(v)$ in \eqref{diff} is $n-K>K-1$. In this situation, the approximate solution that has been obtained can be improved by iteration at the $n-2K+1$ points that are just above $v_1$. This can be done by considering again the constraint equation \eqref{diff} evaluated at the point just below the one that we want to improve, taking into consideration the corrections given by the contribution of the approximate coefficients of the $K-1$ larger nearest points, and (up to) the $K+1$ smaller ones in $\mathcal{L}^{p}_{\varepsilon}$, all of them multiplied by the corresponding functions $f_{k}^{\pm}(v)$.

Finally, it seems natural to impose continuity of the approximate constraint when $v=v_{1}$. From the given construction of the solutions, it is straightforward to check that, for this continuity to be obtained, it suffices to fix the constant $\bar\Lambda_0=\tilde\Lambda(v_1)$, with $\tilde\Lambda(v_1)\gg 1$.

We refer the reader to Appendix \ref{appendixB} for additional details on the suppression of contributions with $\rho \lesssim \rho_m$ in the solutions that we have constructed.

\subsection{Perfect fluids and geometrical corrections}
\label{ivb}

In the previous section we found some approximate solutions to the hybrid Gowdy model that, besides, effectively obey a dynamics dictated by the constraint
\begin{align}
\hat{C}_{\text{app}}=\hat{C}_{\text{FRW}}+\frac{2\pi G \hbar^{2}}{\beta}\left[ e^{2h(\hat{v})}+\hat{O}_{p}\right]H_0
\end{align}
for sufficiently large volumes ($v\geq v_0$). We will now discuss how this constraint can be understood as the one corresponding to an isotropic flat FRW model coupled to different types of perfect fluids coming from the term $e^{2h(\hat{v})}$, and geometrically corrected by the term $\hat{O}_{p}$, which can be  interpreted as arising from homogeneous curvature-like terms or higher-derivative contributions in the gravitational action. Indeed, as discussed in \cite{hybrid-approx}, 
if we choose 
\begin{align}
h(v)=\ln\bigg(\Big[\sum_{w}v^{(1-w)}\Big]^{1/2}\bigg), 
 \end{align}
the dynamics of the constructed states mimics that typical of a content of different perfect fluids with equations of state given by $p=w \epsilon$, where  $p$ and $\epsilon$ denote respectively the pressure and the energy density of the corresponding fluid.
Here, $w$ runs over as many parameters as wanted (one for each different perfect fluid), provided that $w<1$ so that the approximations done in Sec. \ref{sec:app} are valid. This upper bound for $w$ allows for physically interesting couplings such as dust, radiation, and a cosmological constant, obtained by setting $w=0$, $w= 1/3$, and $w=-1$, respectively, in our formulas. Note that the conditions on the peak of the Gaussian-like profiles that are needed for our approximations to hold are then automatically satisfied if $v_0\gg e^{2/(1-w)}$. In addition, we can also treat the case $w=1$ as corresponding to a massless scalar field contribution, coming from the exponential term in the definition \eqref{Op}, that can be included in the homogeneous field $\phi$ of the FRW constraint \cite{hybrid-approx} (just by redefining the latter).

Regarding the new effective term $\hat{O}_{p}$, its action on the kinematical basis provided by $| v\rangle$, given in \eqref{Opact}, can be constructed, for instance, from sums of powers of $\hat{\Omega}^2$, possibly multiplied by smooth functions of $\hat{v}$. Hence, taking into account that the FRW operator $\hat{\Omega}^2$  fully characterizes the curvature scalar $R$ of a flat FRW universe \cite{lqc}, the contribution of $\hat{O}_{p}$ in this modified constraint $\hat{C}_{\text{app}}$ can be seen as a term corresponding to additional curvature-like terms correcting $R$ in the gravitational action. Alternatively, it is also possible to interpret it as discretized higher-derivative terms. These kinds of terms are some of those that one would expect to appear in certain $f(R)$-theories and other modified theories of gravity (see, e.g., \cite{modified}).

To sum up, we have seen how some approximate solutions of the Gowdy model, that is genuinely anisotropic and inhomogeneous,  can effectively behave as solutions (also approximate in general) of the Hamiltonian constraint of a flat FRW model coupled to different types of perfect fluids and with geometrical corrections similar to those of modified gravity. It is worth clarifying that, despite of the dynamical behavior proven for these states with respect to the constraint of the system (namely, a homogeneous and isotropic effective constraint), their Gaussian-like profiles are not peaked on isotropic trajectories of the classical model, but generically on trajectories that are very anisotropic. Classically, isotropy implies the relation $3\Lambda= \ln{(v/2)}$, that ensures that $\lambda_{\theta}=e^{\Lambda}$ coincides with the geometrical average scale factor of the model, given by the cube root of the volume, $v^{1/3}$, up to proportionality factors \cite{awe1}. The Gaussian-like wave functions \eqref{ani-omega}, however, may be peaked on many possible trajectories, determined by quite general functions $\bar{\Lambda}(\hat{\omega})$ of a variety of homogeneous and isotropic operators $\hat{\omega}$. As we have pointed out, in general, these trajectories do not correspond to isotropic and homogeneous solutions of the classical Gowdy system. At the end of the day, it is in the collective behavior of the anisotropies and inhomogeneities, together with the quantum effects of the loop quantization of the geometry, where one finds the ultimate reasons explaining the approximate dynamics of FRW-type, with curvature-like and higher-derivative corrections, that the considered states display. In this sense, the geometrical modifications to the FRW dynamics obtained for these states may be regarded as arising from the underlying quantum theory (both from the loop representation and from the characteristics of the states). Finally, it is worth remarking that such an effective description starts to apply only when one reaches the volume $v_{0}$. For smaller values of $v$ we get instead an effective dynamics which is just that of an FRW model coupled to a homogeneous massless scalar field. Nonetheless, once the epoch with geometrically modified dynamics and perfect fluids is reached, that regime holds indefinitely, for all $v>v_{0}$, by the very construction of our states.

\section{Conclusions}
\label{conclu}

We have investigated the construction of approximate solutions in the hybrid quantum Gowdy model with three-torus topology, linear polarization, LRS, and a minimally coupled massless scalar field \cite{hybrid-matter}. More specifically, we have managed to construct approximate physical states of this inhomogeneous model that in turn are also (approximate) solutions to the Hamiltonian constraint of a homogeneous and isotropic flat FRW model with corrections that can be interpreted as curvature-like or higher-derivative terms. 
The present work significantly generalizes the results of \cite{hybrid-approx}, where we already provided approximate solutions of the Gowdy cosmology that resemble (as far as the Hamiltonian constraint is involved) those of a flat FRW universe with a massless scalar and a perfect fluid. Based on the approximations developed in those previous papers, and extending the analysis carried out there, now we have constructed approximate solutions that behave as those of a geometrically modified flat FRW containing different types of barotropic perfect fluids with equation of state characterized by parameters $w<1$.

Our results show how some specific quantum solutions of inhomogeneous models, in this case the Gowdy $T^3$ model, can behave dynamically as solutions of flat homogeneous and isotropic cosmologies with a particular kind of homogeneous and isotropic matter content, and even with homogeneous and isotropic geometrical modifications that can be regarded as higher powers of the curvature or  higher derivatives. It is worth emphasizing that those solutions are far from being genuinely isotropic and homogeneous. Their anisotropies and inhomogeneities are not negligible, as one could show, in principle, by measuring on those states generic quantum observables beyond homogeneity and isotropy. Despite of that, it is remarkable that those states behave in such a way that they lead to effective terms in the Hamiltonian constraint which are characteristic of a  flat FRW model, and more specifically an FRW universe in the presence of perfect fluids and geometrical corrections. In particular, the dependence of the peak of the considered Gaussian-like profiles on a homogeneous and isotropic geometric operator (of a certain type, though quite general) turns into the appearance of that operator in the effective constraint. This phenomenon, that strongly depends on the specific choice of the considered family of states, emphasizes the fact that effective descriptions generally depend on the particular set of states under analysis. This is an idea which is attracting increasing attention lately (see, e.g., \cite{ale} for discussions on other types of inhomogeneous quantum states for which simple homogeneous descriptions are obtained). In fact, the effective dynamics attained here can be understood to arise from the quantum correlations existing in the considered states between the different sectors of the homogeneous Hilbert space, namely the set of states studied here presents profiles with a specific mixed dependence on the variables of the homogeneous phase space. Besides the mentioned correlations, two key properties of the constructed solutions lay behind this interesting behavior: They display a negligible momentum of the variable that measures the anisotropy and they experience a negligible coupling between the homogeneous sector and the self-interaction of the inhomogeneities. It is because of these properties that one can disregard the most problematic terms in the Hamiltonian constraint, arriving then to a much simpler constraint operator corresponding to a modified flat FRW model coupled to perfect fluids. It is worth mentioning that this approximation is consistent thanks to the quantum geometry effects introduced by LQC, that in particular are responsible of the exponential suppression of the eigenstates of the FRW geometry at small volumes. These quantum effects, together with the collective behavior of anisotropies and inhomogeneities, produce departures from the typical classical behavior predicted by General Relativity in the high-curvature regime, namely, around the cosmological singularity, where the role of the anisotropies would have been very relevant. This is the reason why the effective dynamics obtained for these families of states differ in those regions from the classical solutions of the model.

Our analysis sheds light as well on the generality of our approximations and on the extent to which they are robust. We may expect that these approximations continue to hold as far as one considers states with negligible contributions of the sector $\rho\leq \rho_m$ (with $\rho_m\gg 10$) for the FRW geometry operator, and anisotropies peaked around a certain trajectory on the homogeneous phase space, with a large value of the peak, such that  two more conditions are satisfied. First, the momentum of the anisotropy variable must be small, with a dispersion negligible compared to $\rho_m$, that when squared can be considered a lower bound on the FRW geometry operator $\hat\Omega^2$ acting on our states. Second, the dispersion in the anisotropy variable (dual to that in the momentum of such variable) must be much smaller than the unity. As we explained, the first condition allows us to neglect the anisotropy term, while the second, together with the requirement of a high value of the peak in the anisotropy, permits to neglect the term that contains the self-interaction of the inhomogeneities, and approximate the term with their free contribution using FRW operators. These arguments indicate that Gaussanity on the anisotropies is not strictly necessary. They also tell us that our approximations may be considered stable within a certain sector of quantum states, as long as we do not compromise any of the required properties. 

Another issue that may be helpful in understanding the bases of our approximations is the existence of asymptotic limits in which they arise naturally. That such limits can be found whithin certain schemes follows from the requirements imposed on our states at the beginning of Sec. \ref{sec:sol}. Assuming states peaked on large values of the anisotropy variable, and recalling that, for large $\rho_m=2v_m$, one has $q_\varepsilon \simeq 4/\rho_m$, we can proceed in the following way. First, we focus our attention on functional relations of the form $\sigma_s = q_\varepsilon^\zeta$ with $0<\zeta<1$. For instance, we can take $\zeta=1/2$. Then, the necessary conditions for our approximations are all reached, with improving accuracy, in the asymptotic limit in which $\rho_m$ tends to infinity, in the sector of infinitely large eigenvalues of the FRW geometric operator.

Finally, it may be tempting to extrapolate the lessons learned here about the peculiarities of the effective dynamics associated with certain sets of states in order to build new avenues for the resolution of some of the open questions of the standard cosmological model with inflation. For instance, it may be worthy to investigate whether some elements of its phenomenology, such as the existence of a small but non-vanishing cosmological constant, or the origins of inflation, can be understood in terms of an effective description that arises from a global behavior in the quantum realm, accounting for a multitude of extra degrees of freedom, possibly inhomogeneous and anisotropic in nature. Besides, from a perspective which is beyond that of effective field theories, the results obtained in this work suggest the possibility that some of the corrections to the Einsteinian theory that are nowadays widely investigated in the context of the so-called modified theories of gravity, may actually be rooted in more fundamental quantum geometry effects, occurring in certain types of quantum states.

\acknowledgments
This work was partially supported by the Spanish MICINN/MINECO Project No. FIS2011-30145-C03-02 and its continuation FIS2014-54800-C2-2-P. M. M-B acknowledges financial support from the Netherlands Organisation for Scientific Research (NWO) (Project No. 62001772).

\appendix
\section{Validity of the constraint equation}
\label{appendixA}

For each sector of constant eigenvalue of the Dirac observable $p_{\phi}$ and occupation numbers given by $\mathfrak{n}^{\xi}$ and $ \mathfrak{n}^{\varphi}$ (and hence also with constant eigenvalue of the Dirac observable $H_0$), and for all $\Lambda\in \mathcal L^\varepsilon_{\Lambda^\prime}$, the constraint equation  corresponding to $\hat{C}_{\text{app}}$ can be written on the states \eqref{wave function} as
\begin{align}
\int_{\text{Spc}(\omega)} \text{d}\omega \, N(\omega,p_\phi,\mathfrak{n}^{\xi}, \mathfrak{n}^{\varphi})\hat{C}_{\text{app}}\, e^{-\frac{\sigma_{s}^{2}}{2q_{\varepsilon}^{2}}\left[\Lambda-\bar{\Lambda}(\hat\omega)\right]^{2}}
| \omega\rangle=0.
\end{align}
Let us consider the operator \footnote{Here, we replace square brackets with parentheses in the expressions involving $\Lambda-\bar{\Lambda}$ to avoid confusions with commutators.} 
\begin{align}
\label{commu1}
\bigg[e^{\frac{\sigma_{s}^{2}}{2q_{\varepsilon}^{2}}
\left(\Lambda-\bar{\Lambda}(\hat\omega)\right)^{2}}, \hat{C}_{\text{app}}\bigg]e^{-\frac{\sigma_{s}^{2}}{2q_{\varepsilon}^{2}}\left(\Lambda-\bar{\Lambda}(\hat\omega)\right)^{2}}. 
\end{align} 
Then, we can approximate the constraint by 
\begin{align}\label{b4}
e^{-\frac{\sigma_{s}^{2}}{2q_{\varepsilon}^{2}}\left[\Lambda-\bar{\Lambda}(\hat\omega)\right]^{2}}\sum_{v\in\mathcal{L}^+_{\varepsilon}} N(v,p_\phi,\mathfrak{n}^{\xi}, \mathfrak{n}^{\varphi})\hat{C}_{\text{app}}| v\rangle=0
\end{align}
if we can neglect the contribution of \eqref{commu1} as a correction to $\hat{C}_{\text{app}}$ in this equation. If this is so, the profiles
\begin{align}
N(v,p_\phi,\mathfrak{n}^{\xi}, \mathfrak{n}^{\varphi})= \int_{\text{Spc}(\omega)} \text{d}\omega \, N(\omega,p_\phi,\mathfrak{n}^{\xi}, \mathfrak{n}^{\varphi}) e^{\varepsilon}_{\omega}(v) 
\end{align} 
that satisfy \eqref{const-N} provide indeed approximate solutions to our constraint. 

Let us first recall that, given two operators $\hat{A}$ and $\hat{B}$ whose commutator is negligible in comparison with one of them, the Baker-Campbell-Hausdorff formula implies (see, e.g., \cite{BCH})
\begin{align}\label{bch}
e^{\hat{A}}\hat{B}e^{-\hat{A}}=\hat{B}+[\hat{A},\hat{B}]+\frac{1}{2!}[\hat{A}[\hat{A},\hat{B}]]+\cdots \simeq \hat{B}+[\hat{A},\hat{B}],
\end{align}
where we have ignored higher factor-order contributions, as the commutator $[\hat{A},\hat{B}]$ is assumed to be negligible. Applying this equation, we can rewrite our condition on \eqref{commu1} as the same requirement on
\begin{align}\label{condb7}
\left[\frac{\sigma_{s}^{2}}{2q_{\varepsilon}^{2}}
\left(\Lambda-\bar{\Lambda}(\hat\omega)\right)^{2},\hat{C}_{\text{app}}\right]
\simeq
\frac{\sigma_{s}^{2}}{q_{\varepsilon}^{2}}\left(\Lambda-\bar{\Lambda}(\hat\omega)\right)
\left[\hat{C}_{\text{app}},\bar{\Lambda}(\hat\omega)\right].
\end{align}
On the right-hand side of \eqref{condb7}, we have ignored again higher factor-order contributions.
Taking into account that  the only operator in $\hat{C}_{\text{app}}$ that does not commute with $\bar{\Lambda}(\hat\omega)$ is $\hat{\Omega}^2$, and that, regarding the behavior in $\Lambda$, the Gaussian-like profiles of our states only contribute when ${\sigma_s}[\Lambda-\bar\Lambda(\omega)]/{q_\varepsilon}$ is $\mathcal{O}(1)$, the condition on the right-hand side of \eqref{condb7} (ignoring again factor ordering and irrelevant factors) amounts to demand that 
\begin{align}
\label{condb8}
\frac{\sigma_{s}}{q_{\varepsilon}}e^{-2\bar\Lambda(\hat\omega)}\left[\hat{\Omega}^2,e^{2\bar\Lambda(\hat\omega)}
\right] 
\end{align} 
can be neglected if considered as a correction to the constraint $\hat{C}_{\text{app}}$.

Let us check that this condition is satisfied, and therefore that \eqref{b4} is approximately valid on the studied solutions.
Since, according to \eqref{lamb}, $\bar{\Lambda}(\hat\omega)$ is constant in the region $v<v_{0}$, the corresponding contribution to the commutator appearing in \eqref{condb8} vanishes. Therefore, it suffices to analyze the similar commutator obtained by replacing $\hat{\omega}$ with $\hat{\omega}_0$ in the region $v\geq v_0$. 
Recalling definition \eqref{Op} for $\hat\omega_{0}$, there are two terms to discuss in the commutator: $\left[\hat{\Omega}^2,e^{2h(\hat{v})}\right]$ and $[\hat{\Omega}^2,\hat{O}_{p}]$. As for the first of them, the function $h(v)$ has been chosen to satisfy $h(v\pm 4)\simeq h(v)$ for $v\geq v_{0}$. Therefore, since the action of $\hat{\Omega}^2$, given by \eqref{appOmega2act}, essentially shifts the volume of the state in four units, up and down, it turns out that we can discard the anayzed commutator in our approximations. The second term requires a more careful analysis. On the one hand, the action of this commutator on $| v\rangle$-states with $v>v_{1}+4K$ yields
\begin{align}\label{commu2}
[\hat{\Omega}^2,\hat{O}_{p}]| v\rangle=\sum_{k=0}^{K}&\bigg\lbrace\left[y_{++}(v)f^{+}_{k}(v+4)-y_{++}(v+4k)f^{+}_{k}(v)\right]| v+4(k+1)\rangle\nonumber \\
&+\left[y_{++}(v)f^{-}_{k}(v+4)-y_{++}(v-4k)f^{-}_{k}(v)\right]| v-4(k-1)\rangle\nonumber \\
&+\left[y_{--}(v)f^{+}_{k}(v-4)-y_{--}(v+4k)f^{+}_{k}(v)\right]| v+4(k-1)\rangle\nonumber \\
&+\left[y_{--}(v)f^{-}_{k}(v-4)-y_{--}(v-4k)f^{-}_{k}(v)\right]| v-4(k+1)\rangle\nonumber \\
&+16(kv+2k^2)f^{+}_{k}(v)| v+4k\rangle-16(kv-2k^2)f^{-}_{k}(v)| v-4k\rangle\bigg\rbrace.
\end{align}
These terms [and {\it a fortiori} the corresponding ones in \eqref{condb8}] will give negligible corrections to  $\hat{C}_{\text{app}}$ on our states if $\hat{O}_{p}$ is such that
\begin{itemize}
\item[a)]The functions $f^{\pm}_{k}(v)$ are smooth enough as to satisfy $f^{\pm}_{k}(v\pm 4)\simeq f^{\pm}_{k}(v)$ in the considered region $v\geq v_{0}$.
\item[b)]The integer $K$ is small enough so that $y_{++}(v\pm 4k)\simeq y_{++}(v)$, $y_{--}(v\pm 4k)\simeq y_{--}(v)$, $16kv\ll v^2$, and $32k^2\ll v^2$, for all $k\leq K$ in the sector with $v\geq v_{0}$. Using the fact that, according to their definitions \eqref{appdefini2} and \eqref{appdef2}, the functions $y_{\pm\pm}(v)$ are $\mathcal{O}(v^2)$ in the studied sector, we can say that an integer $K$ would be sufficiently small in this sense if it satisfies $K\ll v_{0}/10$.
\end{itemize}
On the other hand, if one considers the action of the commutator $[\hat{\Omega}^2,\hat{O}_{p}]$ on the remaining sector of $| v\rangle$-states, with $v \leq v_1+4K$, other contributions appear different from those in \eqref{commu2}. This peculiarity occurs because $\hat{\Omega}^2$ is defined on the whole semilattice $\mathcal{L}^+_{\varepsilon}$, whereas $\hat{O}_{p}$ is defined only on the restriction $\mathcal{L}^{p}_{\varepsilon}$. These additional contributions are terms of the form
\begin{align}
\sum_{k=0}^K y_{--}(v_1)f^{-}_{k}(v_1+4k)| v_1-4\rangle, \qquad \sum_{k=0}^{K}y_{++}(v_1-4)f^{+}_{k}(v_1)| v_1+4k\rangle.
\end{align}
Here, the first term accounts for the action of the commutator on all the states $| v_1+4k\rangle$ with $0\leq k\leq K$, and the second one includes the contribution of the action on $| v_1-4\rangle$ (recall that $v_1$ is the lowest point in $\mathcal{L}^{p}_{\varepsilon}$). As a consequence, if conditions a) and b) hold, and 
\begin{itemize}
\item[c)] the functions $f^{-}_{k}(v_1+4k)$ and $f^{+}_{k}(v_1)$ are much smaller than the unit for $0\leq k\leq K$,
\end{itemize}
then all terms under discussion will give negligible corrections to the constraint equation corresponding to $\hat{C}_{\text{app}}$. Let us comment that this last condition c), for $k\neq 0$, was required as well at the end of Subsec \ref{iva} in order to be able to determine the solutions to the approximate constraint. Taking that into account, here we are just including a similar requirement on $f_{0}(v_{1})$.

\section{Suppresion of the solution when \texorpdfstring{$\rho\lesssim\rho_m$}{TEXT}}
\label{appendixB}

In Subsec. \ref{iva} we discussed the construction of approximate solutions to the constraint $\hat{C}_{\text{app}}$. In particular, our analysis contained a region $v<v_0$ where the peak of the Gaussian in the anisotropies is constant. In that region, we made sure that the solution, determined by \eqref{oldsoluA},
possesses only non-negligible contributions for $\rho > \rho_m$ by choosing $\rho_m$ as in \eqref{rhomdef}. One may wonder whether this statement is still true for the whole solution, built for $v\geq v_0$ with Gaussian-like peaks that are not constant anymore.  Although this is not directly granted, because it involves the spectral decomposition of the whole solution in terms of eigenfunctions of the operator $\hat{\Omega}^2$, we show here that the change from \eqref{oldsoluA} to the new solution in  $v\in\mathcal{L}^{p}_{\varepsilon}$ (i.e., for $v\geq v_0$) respects that all relevant contributions have $\rho > \rho_m$, as required.

Up to a global numerical factor, the constraint operator $\hat{C}_{\text{app}}$, given in \eqref{const-app}, can be rewritten in the form $\hat\Gamma-\hat{\Omega}^2$, where
\begin{align}\label{gamm}
\hat\Gamma=\frac{4p_\phi^2}{3\pi G\hbar^2}+\frac{16}{3\beta} e^{2\bar\Lambda(\hat{\omega})}H_0.
\end{align}
For our discussion, the important fact that complicates the analysis in the region $v\geq v_0$ is that the FRW operator $\hat{\Omega}^2$ does not commute with $\hat\Gamma$. Nonethelesss, when acting on solutions to the constraint, the relation $\hat{\Omega}^2=\hat\Gamma$ holds. Besides, for the quantum states that were considered in Subsec. \ref{iva}, characterized by profiles with support on values of the scalar field momentum $p_\phi > p_{\phi}^m$, the action of the operator $\hat\Gamma$ is always greater than (multiplication by) $\rho_m^2$ [see \eqref{rhomdef}], i.e., $(\hat\Gamma-\rho_m^2)$ is strictly positive, because so is the last term in \eqref{gamm}. Hence, on the sector that contains our states, the action of $\hat{\Omega}^2$ will lead only to contributions with $\rho > \rho_m$, as we wanted to show. In fact, one can demonstrate that, on the considered sector, it is a good approximation to neglect the commutator of $\hat{\Omega}^2$ and $\hat\Gamma$, so that it is acceptable to work assuming that they can be simultaneously diagonalized, a fact that supports the conclusion presented above.

\end{document}